\documentclass[a4paper,fleqn,usenatbib]{mnras}

\usepackage{newtxtext,newtxmath}

\usepackage[T1]{fontenc}
\usepackage{ae,aecompl}

\usepackage{graphicx}	
\usepackage{amsmath}	
\usepackage{amssymb}	
\usepackage[usenames,dvipsnames]{xcolor}
\usepackage{soul}

\newcommand{\lp}{LP\,$40$--$365$}
\defcitealias{raddi18}{Paper~I}

\title[Physical and kinematic properties of \lp]{Anatomy of the hyper-runaway star \lp\ with {\em Gaia}}

\author[R. Raddi et al.]{R. Raddi$^{1}$\thanks{E-mail: roberto.raddi@fau.de}, M. A. Hollands$^{2}$,  B. T. G\"ansicke$^{2}$, D. M. Townsley$^{3}$, J. J. Hermes$^{4}$\thanks{Hubble fellow}
\newauthor N. P Gentile Fusillo$^{2}$, D. Koester$^{5}$\\
$^{1}$Dr. Remeis-Sternwarte, Friedrich Alexander Universit\"at Erlangen-N\"urnberg, Sternwartstr. 7, 96049 Bamberg, Germany\\
$^{2}$University of Warwick, Department of Physics, Gibbet Hill Road, Coventry, CV4 7AL, United Kingdom\\
$^{3}$University of Alabama, Department of Physics and Astronomy,  Tuscaloosa, AL, USA\\
$^{4}$University of North Carolina, Department of Physics and Astronomy, Chapel Hill, NC - 27599-3255, US\\
$^{5}$Universit\"at Kiel, Institut f\"ur Theoretische Physik und Astrophysik, 24098, Kiel, Germany}

\date{Accepted 2018 June 6. Received 2018 June 1; in original form 2018 April 25}

\pubyear{2018}

\begin{document}
\label{firstpage}
\pagerange{\pageref{firstpage}--\pageref{lastpage}}
\maketitle

\begin{abstract}
\lp\ (aka GD\,492) is a nearby low-luminosity hyper-runaway star with an extremely unusual atmospheric composition, which has been proposed as the remnant of a white dwarf that survived a subluminous Type~Ia supernova (SN~Ia) in a single-degenerate scenario. Adopting the {\em Gaia} Data Release (DR2) parallax, $\varpi=1.58\pm 0.03$\,mas, we estimate a radius of $0.18\pm0.01$\,R$_{\sun}$, confirming \lp\ as a subluminous  star that is $\simeq 15$ times larger than a typical white dwarf and is compatible with the SN~Ia remnant scenario. We present an updated kinematic analysis, making use of the \textit{Gaia} parallax and proper motion, and confirm that \lp\ is leaving the Milky Way at about $1.5$ times the escape velocity of the Solar neighbourhood with a rest-frame velocity of $852\pm 10$\,km\,s$^{-1}$. Integrating the past trajectories of \lp, we confirm it crossed the Galactic disc $5.0 \pm 0.3$\,Myr ago in the direction of Carina, likely coming from beneath the plane. Finally, we estimate that  \lp\ was ejected from its progenitor binary with a velocity of at least 600\,km\,s$^{-1}$, which is compatible  with theoretical predictions for close binaries containing a white dwarf and a helium-star donor.    
\end{abstract}

\begin{keywords}
stars: individual (GD\,492) 
--- supernova: general --- white dwarfs
--- subdwarfs --- Galaxy: kinematics and dynamics\end{keywords}


%
\section{Introduction} \label{sec:intro}
There is general consensus that Type~Ia supernovae (SN~Ia) are the thermonuclear explosions of white dwarfs \citep{hillebrandt00}. Although a common underlying mechanism makes SNe~Ia  standardisable candles for distances on cosmological scales \citep{riess98,perlmutter99}, their class is rich in peculiar objects, including the subluminous SNe~Iax \citep{foley13}, the calcium-rich transients \citep{perets11}, and SNe~.Ia \citep{bildsten07}. 

Less settled is the discussion on the progenitors of SN~Ia and their close relatives. Most scenarios assume that SN~Ia originate from binary systems, with two fundamentally distinct channels: white dwarfs accreting from a non-degenerate companion (the single-degenerate channel) and mergers of white dwarfs pairs (the double-degenerate channel); for recent reviews, see \citet{wang12,maoz14}.

The chemically peculiar star, \lp\ (aka GD\,492), has been recently proposed by \citet{vennes17} as a partially burned white that survived a SNe~Iax explosion \citep[see][]{jordan12,kromer13,kromer15}. The detection of a significantly super-Solar manganese-to-iron ratio \citep[][hereafter Paper~I]{raddi18} suggests that \lp\ had a non-degenerate companion \citep{seitenzahl13a,cescutti17}. In the proposed scenario, \lp\ was unbound from the original binary and, due to its initially large orbital speed, it is now travelling at more than 500\,km\,s$^{-1}$ (corresponding to the measured radial velocity), becoming therefore a runaway star\footnote{Runaway stars are proposed to gain their momentum via ejection from binary SN explosions \citep{portegies-zwart00}. Hyper-runaway stars  are the fastest of this  class, with velocities comparable to those of traditional hypervelocity stars, which are thought to form via multi-body interactions with super-massive black holes \citep{hills88,brown15}.}. 

In this Letter, we use the accurate parallax and proper motions available from the recent {\em Gaia} Data Release~2 \citep[DR2;][]{gaia16,gaia18} to carry out the first detailed kinematic analysis of \lp, which provides strong constraints on its physical parameters, its past trajectory, and the properties of the progenitor system.

\section{Physical properties} \label{sec:one}
\begin{figure}
\includegraphics[width=\linewidth]{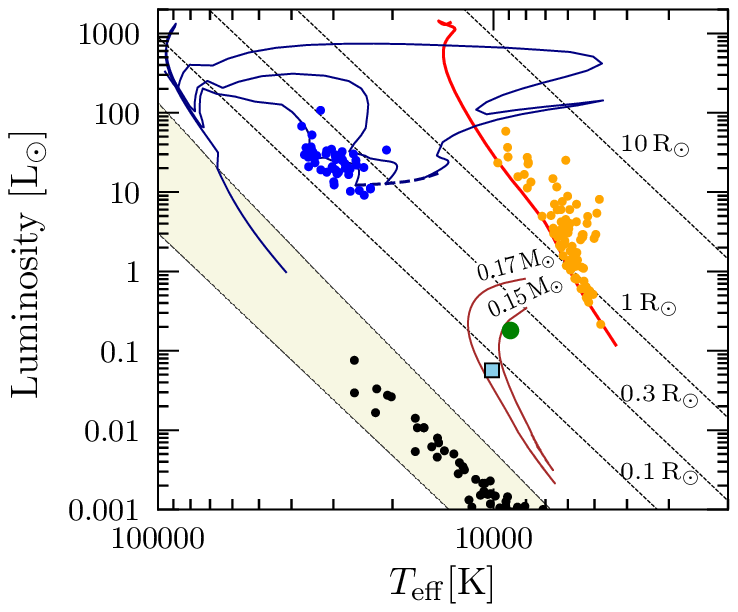}
\caption{Hertzsprung-Russel diagram displaying \lp, compared to various classes of stars. We plot the parameters based on our analysis including the {\em Gaia} data (green circle) and the  \citet{vennes17} estimate (light-blue square), hot subdwarfs \citep[blue dots;][]{lisker05}, main sequence stars \citep[orange dots;][]{boyajian13}, and nearby white dwarfs \citep[black dots;][]{giammichele12}. The proto- white dwarf sequence and the cooling tracks \citep{althaus13} are shown for 0.15 and 0.17~M$_{\sun}$ white dwarfs (brown solid curves). For reference, we draw the canonical-mass white dwarf cooling sequence \citep[beige-coloured strip;][]{fontaine01},  the 100~Myr old main sequence \citep[red curve;][]{choi16},  and the evolutionary tracks for 0.47, 0.48, and 0.50~M$_{\sun}$ hot subdwarfs \citep[blue curves;][]{dorman93}. The  luminosities for given stellar radii are shown as a function of $T_{\rm eff}$  (dotted curves).\label{f:hr}}
\end{figure}
\begin{table}
\caption{Physical parameters of \lp. The nominal values of $d$, $R$ and $M$ correspond to the median of the distributions, and the 1 $\sigma$ uncertainty. The 5--95 per cent range is also given below.\label{t:phys}}
\centering
\scriptsize
\begin{tabular}{@{}cccccc@{}}
\hline
$T_{\rm eff}^1$ & $\log{g}^1$ & $d$ & $R$ & $M$ & $L$ \\
\hline
(K) & (cgs) & (pc) & (R$_{\sun}$) & (M$_{\sun}$) & (L$_{\sun}$)\\
\hline
$8900 \pm 300$ & $5.50 \pm 0.25$ & $632 \pm 14$ & $0.18 \pm 0.01$ & $0.37^{+0.29}_{-0.17}$ & $0.18 \pm 0.01$ \\
& & $610$--$655$& $0.16$--$0.20$& $0.14$--$0.98$ & $0.17$--$0.19$  \\
\hline
\multicolumn{2}{l}{$^1$From \citetalias{raddi18}.}
\end{tabular}
\end{table}

The precision of the {\em Gaia} parallax of \lp, $\varpi = 1.58 \pm 0.03$\,mas, ensures a direct conversion between parallax and distance without significant loss of accuracy \citep[][and references therein]{bailer-jones18}, placing \lp\ at  $632 \pm 14$\,pc. 

Scaling our best-fit model from \citetalias{raddi18} to the {\em Gaia} magnitude, $G_{\rm p} = 15.58$\,mag, which we corrected for the sightline interstellar extinction \citep[$0.02$\,mag;][]{green18}, we estimate the integrated flux density of \lp\ to be $f = 1.45 \times 10^{-11}$~erg/cm$^2$/s. Thus, using the \textit{Gaia} parallax and the Stefan-Boltzmann law, the radius of \lp\ is constrained to $R = 0.18 \pm 0.01$R\,$_{\sun}$, accounting for the parallax and $T_{\rm eff}$ uncertainties. This result contrasts with the estimate of 0.07\,R$_{\sun}$, obtained by \cite{vennes17} via interpolation of their $T_{\rm eff}$ and $\log{g}$ with cooling models for low-mass helium-core white dwarfs \citep[][]{althaus13}. 
On the Hertzsprung-Russel diagram (Fig.\,\ref{f:hr}), our new results place \lp\ at a cooler and brighter location with respect to the parameters of \citet{vennes17}. 

Combining our radius estimate with the surface gravity derived from the spectral fit in \citetalias{raddi18} implies the mass of \lp\ is constrained as $M \propto g R^{2} = 0.37^{+0.29}_{-0.17}$\,M$_{\sun}$, with the 5--95 per cent confidence range between $M = 0.14$--0.98\,M$_{\sun}$. We note  that \lp\ does  not match the radius-luminosity relation of main sequence stars, as it is two orders  of magnitude  less luminous than stars of similar $T_{\rm eff}$ (A-type stars), while it is hotter than main-sequence stars of similar radii (M-type dwarfs). \lp\ also diverges from the mass-radius relation of both canonical white dwarfs \citep{tremblay17} and low-mass helium-core white dwarfs \citep{althaus13}, with a composition clearly excluding a membership to the latter class  of stars.

\begin{figure}
\includegraphics[width=\linewidth]{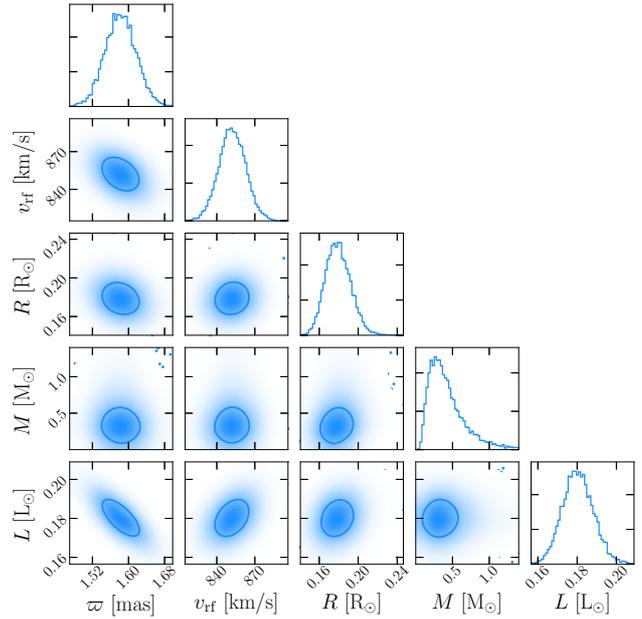}
\caption{Corner plot displaying the correlation between the {\em Gaia}-DR2 parallax and the physical properties of \lp. The 1 $\sigma$ contours are shown on the correlation plots.\label{f:mcmc}}
\end{figure}

We will discuss the present appearance of \lp\ with reference to its evolutionary status in Section~\ref{sec:four}. The physical parameters are summarised in Table\,\ref{t:phys}, while their correlation with $\varpi$ is shown in Fig.\,\ref{f:mcmc}.

\section{Kinematic analysis} \label{sec:two}
\begin{figure*}
\includegraphics[width=\linewidth]{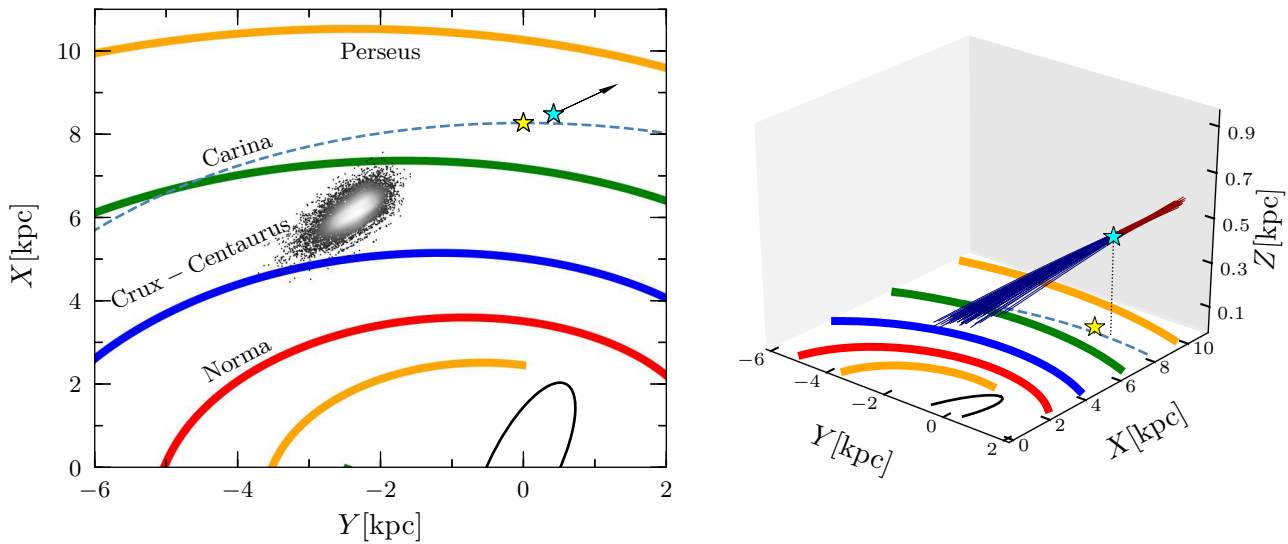}
\caption{Simulated trajectories of \lp\ in the Milky Way potential. In both panels, the Sun and \lp\ are represented as a yellow and a cyan star, respectively, while the spiral arms and the Galactic bar are shown as colored and black curves \citep[][]{vallee08}, respectively. The Solar circle is traced by a dashed curve. {\em Left panel:} The reference frame follows the standard left-handed Galactocentric convention, with the Galactic centre at $X,Y = (0,0)$ and the $Y$-axis oriented in the direction of the Galactic rotation. The arrow shows the mean escape direction of \lp. The cloud of dots represents the $X$-$Y$ coordinates of the trajectories crossing the Galactic plane ($Z = 0$). {\em Right panel:} Three-dimensional representation of a random sample  of simulated trajectories. Past and  future trajectories are plotted as dark-blue and red curves, respectively. The vertical dotted line shows the distance $Z$ of \lp\ from the plane. Note the scale on the $Z$ axis is expanded  with respect to the $X$-$Y$ plane to enhance the visualization.\label{f:tr}}
\end{figure*}

The high precision of {\em Gaia} parallaxes enables measuring the total rest-frame velocity of \lp\ with no {\em a priori} assumption. Taking into account the correlation between the astrometric quantities, and using the radial velocity from \citetalias{raddi18}, $v_{\rm rad} = 499 \pm 6$\,km\,s$^{-1}$, with the Galactic parameters described  below, we estimate the rest-frame velocity as $v_{\rm rf} = 852 \pm 10$\,km\,s$^{-1}$, making it the fastest known hyper-runaway star that is the nearest to the Sun. At this remarkably large speed, \lp\ exceeds the Galactic escape velocity \citep[$520$--$533$~km\,s$^{-1}$ in the Solar neighbourhood;][or $\approx 550$\,km\,s$^{-1}$ as for the adopted Galactic model]{piffl14,williams17}, thus it is confirmed as gravitationally unbound from the Milky Way.

To investigate the space motion of \lp, we used the Galactic orbit integrator implemented in the {\sc python} package {\sc galpy}\footnote{\url{https://github.com/jobovy/galpy}} \citep{bovy15}. We set up the Galactic potential described in \citet{bovy13}, which consists of three components (bulge, disk, and halo), to which we added a central black hole of $4 \times 10^{6}$\,M$_{\sun}$ \citep{bovy15}. In this model, following \citet{schonrich12}, the Sun is placed at $R_{0} = 8.27 \pm 0.29$\,kpc from the Galactic centre and the Milky Way rotation speed at the Solar circle is $V_{\rm{c}} = 238 \pm  9$~km\,s$^{-1}$,  while the peculiar motion of the Sun in the Local Standard of Rest is $(U_{\sun},\,V_{\sun},\,W_{\sun}) =  (11.1,\, 12.24,\, 7.25)$~km\,s$^{-1}$, from \citet{schondrich10}.  Our simulations take into account the statistical and systematic uncertainties quoted in the original works. The boundary conditions correspond to the {\em Gaia} DR2 observables, $\alpha$, $\delta$, $\mu_\alpha$, $\mu_\delta$, and $\varpi$, along with $v_{\rm rad}$ from \citetalias{raddi18}. We sampled the boundary conditions via a Monte Carlo method, taking into account the {\em Gaia} covariance matrix and assuming Gaussian distributions for the Galactic model parameters. We back-traced the trajectories for 250\,Myr in the past, i.e. the timescale of one Solar orbit around the Milky Way at the Galatocentric distance of the Sun. We display a representative set of trajectories in Fig.\,\ref{f:tr}.

From this simulation, we find that \lp\ crossed the Galactic plane in the inter-arm region between the Crux-Centaurus and the Carina spiral arms, at $4.2 \pm 0.5$\,kpc from the present position of Sun, corresponding to a Galactocentric radius $R_{\rm G} = 6.0 \pm 0.3$\,kpc. At its speed, \lp\ has travelled for $5.0\pm 0.3$\,Myr to reach its current location after it crossed the Galactic disc. We do not attempt to associate \lp\ to any  known Galactic  structure  along its trajectory (e.g. clusters, spiral arms, or stellar streams), given that its age is not sufficiently well constrained. However, we note that the flight time from a Galactocentric  distance of  100\,kpc ($Z = -7.3 \pm 0.5 $\,kpc) is $\approx 140$\,Myr, setting an upper limit on the cooling age, if \lp\ was ejected from the Galactic halo. We speculate that its membership to known structures could be investigated in future, when evolutionary models will become available. An origin in Galactic centre or in the Magellanic Clouds can definitely be excluded.  

\begin{figure}
\includegraphics[width=\linewidth]{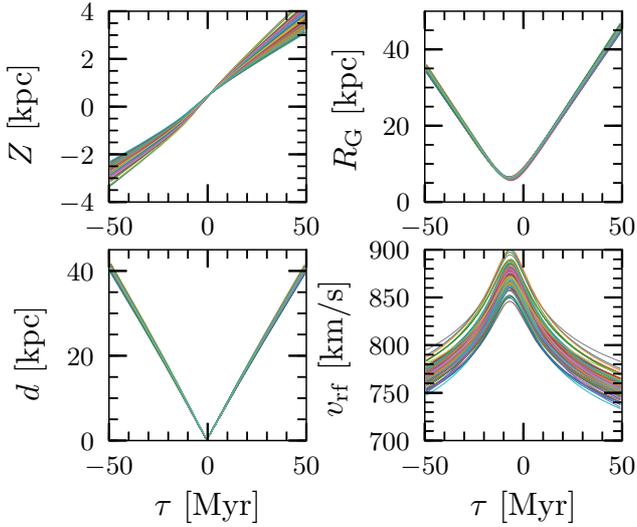}
\caption{Evolution of positional and kinematic parameters for a representative sample of \lp\ trajectories, plotted as function of the flight time, $\tau$. Clockwise from top left: Galactic $Z$ coordinate; Galactocentric radius $R_\mathrm{G}$; rest frame velocity $v_\mathrm{rf}$; Heliocentric distance $d$. $\tau = 0$ corresponds to the {\em Gaia} DR2 epoch. \label{f:kin}}
\end{figure}

In Fig.\,\ref{f:kin}, we show the evolution of positional and kinematic parameters of \lp. We note the asymptotic behaviour of $Z$, $R_{\rm G}$, and $d$, which is typical of open trajectories. We also note that \lp\ was accelerated while crossing the Galactic disc $\approx 5$\,Myr ago and it is slowing down while leaving the Milky Way. All the future trajectories reach  $R_{\rm G} = 100$\,kpc (not plotted in Fig.\,\ref{f:kin}) in $\approx 130$\,Myr.

The simulated trajectories have small pitch angles with respect to the plane, $\gamma\,\approx 5.6$\,deg, potentially implying a boost from the Galactic rotation \citep[see][]{kenyon14} if the ejection occurred in the Galactic disc. The ejection velocity from the putative progenitor system, $v_{\rm ej}$, could be estimated from $v_{\rm rf}$ via:
\begin{equation}
v_{\rm rf}^{2} = v_{\rm ej}^{2} + V_{\rm c}^{2} + 2\,v_{\rm ej}\,V_{\rm c}\,\cos{\beta}\,\cos{\gamma},
\label{eq:vrf}
\end{equation}
where $v_{\rm c}$ is the rotational velocity of the progenitor in the Milky Way at the moment of the explosion, and $\gamma$  is the angle between $v_{\rm c}$ and $v_{\rm ej}$ in the Galactic  plane. However, we stress that Eq.\,\ref{eq:vrf} only holds if \lp's progenitor exploded  close  to the  Galactic  plane. In this case, we would have that $v_{\rm ej} \sim 600$\,km\,s$^{-1}$. Given that evolutionary timescales of a peculiar white dwarfs such as \lp\ are yet unconstrained, it cannot be excluded with certainty that the progenitor exploded at several kiloparsecs from the Galactic centre. Hence, the contribution to $v_{\rm rf}$ due to the Galactic rotation would be negligible for a halo progenitor, so that $v_{\rm ej} \approx v_{\rm rf}$.

\section{Properties of the binary progenitor} \label{sec:four}
Given that  \lp\ is a single star \citep[][]{vennes17,raddi18}, if it is the partly-burnt remnant of a SN~Iax, the progenitor system must have been separated as consequence of mass loss caused by the SN explosion. 

Following \citet{hills83}, under the assumption of instantaneous mass loss, a binary system breaks apart if it experiences a minimum mass loss, $\delta M$, defined as:
\begin{equation}
\left\langle \bigg(\frac{\delta M}{M_{\rm prog} + M_{\rm donor}}\bigg)_{\rm min} \right\rangle\,= 0.5 \times \bigg[ 1 - \bigg(\frac{v_{\rm kick}}{v_{\rm orbit}}\bigg)^{2} \bigg],\label{eq:eq2}
\end{equation}
where $M_{\rm prog}$ and $M_{\rm donor}$ are the mass of progenitor and donor stars, respectively, and $v_{\rm kick}$ is a velocity kick that could arise from e.g.\ asymmetric explosions \citep{jordan12}. 

In \citetalias{raddi18}, we brought evidence in favour of a single-degenerate scenario via the detection of a super-Solar manganese-to-iron ratio, which requires a $M_{\rm prog} > 1.2$\,M$_{\sun}$ to synthesise manganese \citep[][]{seitenzahl13a}. Thus, considering the simplest case, with negligible  mass-loss from the donor star \citep[as shown by][for non-giant companions]{marietta00,pan12,liu12,liu13} and no birth kick ($v_{\rm kick} = 0$), we have that the minimum possible mass of the  donor star is $M_{\rm donor} = 0.86^{+0.34}_{-0.58}$\,M$_{\sun}$. Hence, the donor star can be constrained to $\leq 1.32$\,M$_{\sun}$ at the 95 per~cent confidence  limit.  Numerical simulations of subluminous SN explosions that form partly-burnt remnants also produce $v_{\rm kick}$ of up to a few hundred km\,s$^{-1}$, although debated \citep{jordan12,kromer13}, that would enable to unbind binaries with even more massive donors. 

Considering possible ejection velocities at the moment of explosion, $v_{\rm ej} \lesssim v_{\rm rf} = 600$--$800$\,km\,s$^{-1}$, the orbital period of a near-Chandrasekhar mass white dwarf with a donor star between $0.3$--$1.2$\,M$_{\sun}$ is in the range of $30$\,min to 1\,hr.
Binaries leading to such tight configurations have been theoretically identified in the helium star donor channel of the single-degenerate scenario \citep{wang09c,wang14}. These systems are proposed as a leading mechanism behind subluminous SNe~Iax \citep{jha14,mccully14}, which could also have an important contribution to the population of Galactic hyper-runaway stars \citep[][]{justham09,wang09b}. They include either massive helium core- or shell-burning donor stars (up to $2.5$\,M$_{\sun}$),  down to canonical mass hot subdwarfs \citep[$\approx 0.47$\,M$_{\sun}$;][]{heber16}.  As already noted by \citet{vennes17}, the hypervelocity hot subdwarf US\,708 \citep{geier15} and \lp\ could represent two sides of the same coin: a donor star and an unburnt remnant surviving a SN~Ia in a close binary.  US\,708 is suggested to have a mass of $0.3$\,M$_{\sun}$, and it is proposed to have originated as the donor in an ultra compact 10-min binary exploded as sub-Chandrasekhar mass SN. \lp, instead, is unique in its kind, as it is the first surviving white dwarf fitting with this scenario. Gaining a speed as large as that of of US\,708 \citep[$\sim 1200$\,km\,s$^{-1}$;][]{geier15}, the subdwarf donor of \lp\ could have escaped the Milky Way long time ago. 

However, we note that the hot subdwarf donor scenario may not entirely fit with \lp, because low-mass donors ($\lesssim0.8$\,M$_{\sun}$) are not expected to lead to near-Chandrasekhar mass SN~Ie, due to unstable helium-shell burning \citep{piersanti14,brooks15}. More massive donors \citep[$0.8$--$1.7$\,M$_{\sun}$;][]{piersanti14,brooks17}, instead, could lead to near-Chandrasekhar mass explosions, after accreting sufficient mass on to the white dwarf. While such progenitors could be identifiable as helium-novae, or super soft X-ray sources, some detached candidates of this SN channel have been proposed, e.g.\ KPD\,1930+2752 \citep{maxted00,geier07}. Although this scenario is theoretically appealing, very compact systems like these remain, so far, rare \citep[just three are known with $P_{\rm orb} < 90$\,min;][]{vennes12,kupfer17b,kupfer17a} and their fate is still debated. 

In this context,  \citet{shen18} have recently identified three hyper-runaway stars classified as ``expanded'' white dwarfs, which are proposed as former donor stars in double-degenerate systems. These stars would now  have larger luminosities than normal  white dwarfs due to several effects, such as dynamical interactions within the progenitor binary, impact of the SN ejecta, and accretion of radioactive $^{56}$Ni-rich material, all of which contributed to (partially) lift the core degeneracy and expand the envelope. Although having larger $v_{\rm rf}$, redder colours, and a  likely carbon-oxygen atmosphere, \citet{shen18} proposed  these new stars and \lp\ to form a short-lived heterogeneous class of stars that would eventually re-join the white dwarf cooling  sequence. As the peculiar atmospheric composition of \lp\ remains the strongest connection to the  partly-burned  remnants  of SN~Iax, we note that such objects would also expand due to internal adjustments of the stellar structure after the explosion \citep{kromer13,shen17}. Further theoretical investigations concerning the evolutionary timescales are crucial, given the small parameter space tested in the literature.

Finally, another physical constraint worth comparing is the rotational velocity of \lp\ to that of US\,708. The hot subdwarf shows a high rotational velocity \citep[115~km\,s$^{-1}$;][]{geier15}, consistent with it being tidally locked at an orbital period before detonation of roughly 10 min.  US\,708 is best explained as the donor to an exploded system that has not significantly evolved in size since the event.

On the other hand, we do not measure any rotational broadening for \lp, and can only establish an upper limit on the rotational velocity of 50~km\,s$^{-1}$ (\citetalias{raddi18}). To have been ejected from the binary with $v_{\rm ej} > 600$~km\,s$^{-1}$ it is likely that \lp\ became unbound from a tight system with an orbital period shorter than 1 hr. Tides are likely to synchronize the rotation of both components of such short orbital periods \citep{fuller12}.

The rotation velocity of a 0.18\,R$_{\sun}$ star rotating faster than 1 hr would exceed 200~km\,s$^{-1}$, which we would be able to detect in \lp\ for all inclinations higher than roughly 20 deg. However, if the progenitor of \lp\ were a massive white dwarf rotating at the previous orbital period, its radius will have increased more than a factor of 20 to match the size we observe today; its rotation would have likely slowed considerably to conserve angular momentum. Therefore the relatively slow rotation velocity adds another line of evidence to \lp\ being the partly burnt remnant of a sub-luminous supernova and not the donor.

\section{Summary and conclusions} \label{sec:seven}
\lp\ is a hyper-runaway star, which is the first example of a possible SN~Iax survivor, i.e. a partly burnt white dwarf that is enriched with nuclear ashes \citep{vennes17,raddi18}. Here, we have presented a detailed analysis of its physical and kinematic properties, making use of the recent {\em Gaia}-DR2 astrometry. At a distance of $632 \pm 14$\,pc, \lp\ is the nearest hyper-runaway star to the Sun that is unbound from the Galaxy, having a rest-frame velocity of $852 \pm 10$\,km\,s$^{-1}$. We confirm it as a subluminous star, with a radius of $0.18 \pm 0.01$\,R$_{\sun}$ and a mass of $0.37^{+0.29}_{-0.17}$\,M$_{\sun}$, matching the partly-burnt white dwarf hypothesis. 

We simulated the past trajectory of \lp\ in the Milky Way potential, finding it crossed the Galactic disc $\approx 5$\,Myr ago in the direction of Carina at a Galactocentric radius  of $\approx 6.5$\,kpc, i.e. $\approx 4.2$\,kpc  from the Sun's current position. With a cooling age possibly as large as $\sim 100$\,Myr, \lp\ could have been ejected from the Galactic halo at a few kpc below the plane. From the constraint we have on \lp's mass, we suggest its progenitor may have been a short-period ($30$--$60$\,min) binary with a donor star of $0.8$--$1.32$\,M$_{\sun}$.  The ejection velocities from such tight binaries are sufficient to  accelerate \lp\ and similar objects to and beyond the Galactic escape velocity.

With the high quality data of {\em Gaia} DR2, we expect that more candidates of partially burned remnants and/or donor stars ejected from SN~Ia will be identified, helping to fill in the gaps between theoretical predictions and unusual stars such as \lp\ and those found by \citet{shen18}.

\section*{Acknowledgements}
We thank U. Heber for useful discussions. R.R. acknowledges funding by the German Science foundation (DFG) through grants HE1356/71-1 and IR190/1-1. The research leading to these results has received funding from the European Research Council under the European Union's Seventh Framework Programme (FP/2007-2013) / ERC Grant Agreement n. 320964 (WDTracer). Support for this work was provided by NASA through Hubble Fellowship grant \#HST-HF2-51357.001-A, awarded by the Space Telescope Science Institute, which is operated by the Association of Universities for Research in Astronomy, Incorporated, under NASA contract NAS5-26555. 

This work has made use of data from the European Space Agency (ESA) mission {\it Gaia} (\url{https://www.cosmos.esa.int/gaia}), processed by the {\it Gaia} Data Processing and Analysis Consortium (DPAC, \url{https://www.cosmos.esa.int/web/gaia/dpac/consortium}). Funding for the DPAC has been provided by national institutions, in particular the institutions
participating in the {\it Gaia} Multilateral Agreement.




\bibliographystyle{mnras}






\bsp	
\label{lastpage}
\end{document}